# WEB-BASED DATABASE MANAGEMENT TO SUPPORT TELEMEDICINE SYSTEM

Hafez Fouad

Microelectronics Dept., Electronics Research Institute, Cairo, Egypt

## ABSTRACT

*The transfer of the medical care services to the patient, rather than the transport of the patient to the medical services providers is aim of the project. This is achieved by using web-based applications including Modern Medical Informatics Services which is easier, faster and less expensive. The required system implements the suitable informatics and electronics solutions efficiently for the Tele-medicine care. We proposed an approach to manage different multimedia medical databases in the telemedicine system. In order to be efficiently and effectively manage, search, and display database information, we define an information package for both of doctor and patient as a concise data set of their medical information from each visit. The methodology for accessing various types of medical records will be provided, also we will design two web-based interfaces, high-quality data and display for many medical service purposes.*

## KEYWORDS

*Telemedicine, Medical database, Teleconferencing, Teleconsultation, Telediagnos is, Medical Informatics, web-based medicals applications.*

## 1. INTRODUCTION

There are shortages of medical resources in rural areas or geographically isolated regions, so many physicians may be reluctant to serve in these areas. Therefore, people who live there will receive lower medical care than those who live in urban areas. There is an important need to develop a telemedicine system to improve the quality of medical services there and provide more educational opportunities to the physicians in these areas [1]–[4].Telemedicine can be defined as the providing of medical services over a distance. The Archiving and Communication System (PACS) will be used in the telemedicine process as this service requires patient history, medical images, and related information. By using PACS [5]–[11], we can find that the integrated telemedicine system consists of the following five subsystems:

1) Acquisition subsystem;
2) Viewing subsystem;
3) Teleconferencing subsystem;
4) Communication subsystem;
5) Database management subsystem.

The first subsystem is the acquisition subsystem which collects multimedia information [12] then converts it to a standard format (e.g., DICOM 3.0 [13]). The second one is the viewing subsystem which displays and manipulates the images and other medical information [14]–[15]. The third one is the teleconferencing subsystem which allows face-to-face interactive conference between physicians in rural areas and medical centers [16]–[18],this subsystem is not included in a PACS. The forth one is the communication subsystem which includes the connectivity method; local area networks (LAN's)and a wide area network (WAN) to transmit and receive data[19]–[21].The patient medical record consists of the patient complaint, history of illness, results of physical examination, laboratory tests, and diagnostic images. The medical information may be of the following types: text, voice, image [e.g., x-ray, computed tomography (CT), or magnetic resonance imaging (MRI)], and dynamic video (e.g., videoesophagogram and endoscopy) [22]–[24]. Thus, it is essential to design a medical information database for managing a huge amount of heterogeneous data. In some studies [14],[25]–[27] However, this approach may complicate archiving operations and introduce an inconsistency problem while concurrently accessing the image data [28]–[30]. This management approach may make it difficult to access the videotapes and share themsimultaneously. Moreover, the integration of video with text and images in a telemedicine system is a problem.

To solve these problems, a data management methodology is proposed which is the fifth subsystem, by which medical information can be organizedbased on the patient's complaint as well as the medical history. This will supporta unified interface for manipulating and accessing the different types of all medical information mentioned above. The management of medical databases and the user interface has been implemented as major components of a telemedicine system through A in Medical. Com web-Portal.

# 2. SYSTEM ANALYSIS AND DESIGN

## 2.1. Telemedicine System Service

In this paper, we have developed a telemedicine system that supports teleconsultation, telediagnosis, and tele-education. In *teleconsultation*, rural physicians referred their patients to the medical specialists at a medical center who provide second opinion for them. The patient's medical records will be shared between the rural physicians and the specialists; they will discuss the symptoms of the patient's conditions interactively. The patient's final diagnosis is reached following discussion between the two physicians.

In *teleconsultation*, we need a synchronous two-way videoconferencing system as well as a document-sharing mechanism to allow rural physicians to send their patient's medical information to specialists and engage in face-to-face conversation. In *telediagnosis*, it is similar to *teleconsultation*, but the specialist makes a diagnosis based on the received information. The specialist makes the diagnosis and then forwards the diagnosis report to the rural physician. The major difference between them is that the *telediagnosis* requires high-quality data and images to achieve an accurate diagnosis, while the *teleconsultation* requires a synchronously interactive conference environment. Telediagnosis can be performed asynchronously. In *tele-education*, a rural physician playing a student role obtains advanced medical expertise from the specialists. There are two ways to deliver tele-education to rural physicians. First, knowledge may be delivered in a face-to-face manner through teleconferencing between the rural physician and the specialist. So, a real-time videoconferencing system capability is required for interactive communication. Second, the knowledge may be put in medical teaching materials which can be organized and converted to a digital multimedia textbook presented on the World Wide Web (WWW). A network discussion panel may also be created for exchanging ideas and discussing problems among the rural physician and the specialist. Rural physicians can access these materials and educate themselves via the Internet. So, an authoring tool for compiling the medical teaching materials and a friendly userinterface for browsing and discussing the multimedia textbook are required.

In order to meet the requirements of *teleconsultation, telediagnosis, and tele-education* simultaneously, patient medical records and the associated images must be organized in such a way that a physician can easily access the database based either on a patient's clinical history or on particular cases (clinical problems). This requires that the database must meet different purposes by providing both patient-oriented data folders and problem-oriented data folders. A patient-oriented data folder is used to store all the medical records of a single patient; a problem-oriented data folder is used to store all the medical records of one specific case.

## 2.2. Conceptual Databases Models

We know that the physician makes a diagnosis and treatment plan in the clinical practice not only based on the patient's current situation, but also on a review of the patient's history and references in similar disease symptoms. The current traditional medical databases are constructed according to the type of material in the records. These records may be laboratory data, consultant comments, physicians' notes, and diagnostic medical images from different sources and each of them were managed in separate files. Although this management method is relatively easy to maintain, it is difficult to trace the history of particular problem. To resolve this difficulty, we defined a database as a concise data set containing all of the medical diagnostic information of the patient. Besides, the database package can manage and save any change of status or new information that emerges from the subjective description, objective description, assessment, and plan; these derivations are based on subjective, objective, assessment, and plan (SOAP) medical record methodology [30].

The subjective description (S) refers to the description of a patient's chief complaint and the history of the disease problem. It is interpreted from the patient's point of view, and in **this** study, includes symptom code, duration, location, severity, description, and chief complaint. The objective description (O) records the results of all measurements during the current visit and factual plan results as noted by the physician during the previous visit concerning the same problem. In this part, physical examination results, laboratory data, and diagnostic plan conclusions are summarized in the fields of item, location, finding, sign-code, and description. The assessment information, part A, records the physician's diagnosis and a description of the disease problem based on the information in part S and part O It is expressed with the problem ID and an assessment description. The plan information, part Prefers to the diagnostic and therapeutic plans made by the physician specifically addressing the patient's problem.

## 2.3. Database Implementation

It is noteworthy that, as in Fig. 1, part of the medical record is a form of multimedia. An important point in system design is how to build a medical information database system to manage heterogeneous data. Although the relational database provides a set of powerful tools to manipulate data, its template of predefined data type limits its ability to manage large objects. In our implementation, the attributes of Video and Images are defined as FILE type. The attributes of Report, Chief Complaint, Description and other attributes are defined as TEXT type. More importantly, they can be uniformly manipulated in SQL queries.

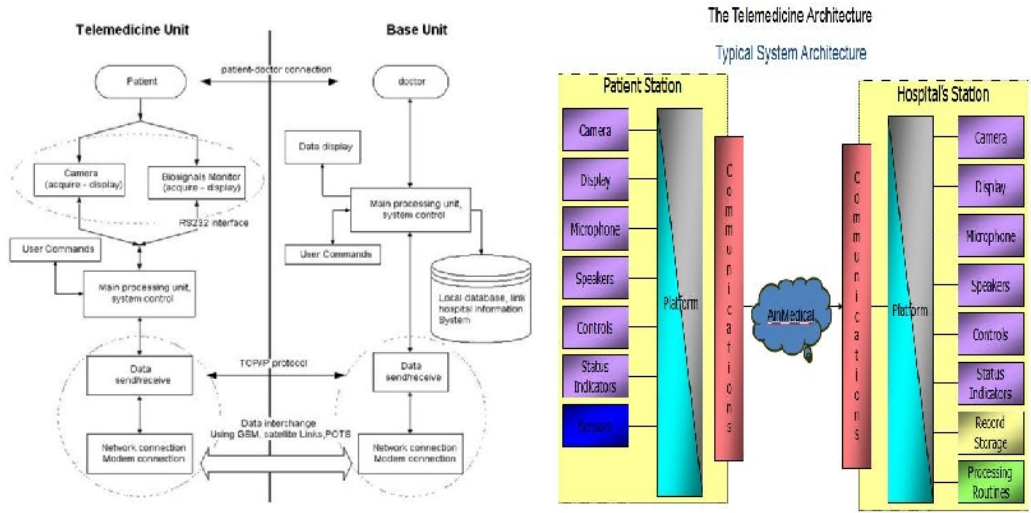

Figure 1.Web-based telemedicine system Arch.[1]Figure 2.Ain Medical Telemedicine Architecture

In addition to data integration, speed of data retrieval is also a factor that affects the performance of the telemedicine system. In this paper, a three-layer hierarchical database is created; the three layers consist of main database, long-term database, and local database. The main database stores medical information concerning patients who have visited within recent months. After this period, the data are moved to a long-term database. Then the long-term database server packs the image data according to time of creation and manages it in the DICOM media storage directory (DICOMDIR) format, which is introduced by the American College of Radiology and the National Electrical Manufacturers Association (ACR/NEMA)to store DICOM-formatted medical images in permanent media[31].

The local database provides a short-term storage location forthe medical records of patients currently visiting. It functions to reduce workload of the database server and traffic of the network. In order to prepare the most frequently used data, the PREFETCH mechanism, which works to reduce the data accessing time, is incorporated into the local database installed in the medical center. During teleconsultation, the PREFETCH precedes the diagnosis and accesses medical records according to the schedule. In telediagnosis, the medical records must also be prefetched if the diagnosis report has not yet been completed. Moreover, the REFRESH mechanism is also incorporated in the local database at the rural site to maintain acceptable communication reliability. It stores the medical records of newly visiting patients in the local database and forward these records when the communication channel has been successfully connected. Thus, it can avoid data loss caused by failure of the communication channel.

## 2.4. AinMedical.com Database

1)*Doctor Registration*, for the doctor to become a member of the A in Medical portal, the system requires a registration of a new user as Doctor which allows him to access the AinMedical.com web- portal different services. The Doctor who uses website with Pre-condition having a valid email address to complete registration. A in Medical portal sends message to new doctor to activate his account. Also a message will be sent to A in Medical's Administrator to approve the new doctor account or delete it. If A in Medical's Administrator approved the doctor registration, A in Medical presents welcome page for New Doctor and provide link to login his account. As indicated in Fig. (3).

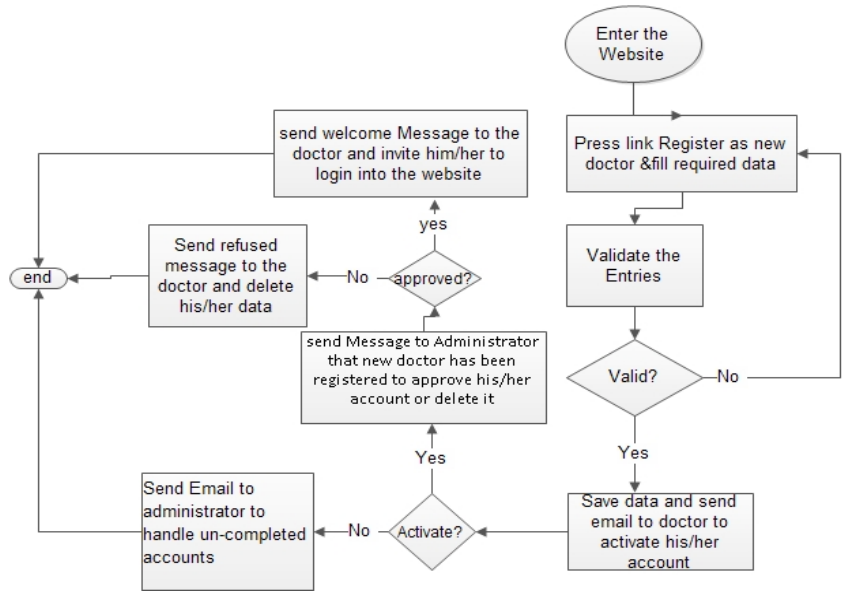

Figure 3.Doctor Registration

**2)** ***Patient Registration***, for the patient to become a member of the AinMedical.com web-portal, the system requires a registration of a new user as Patient which allows him to access the A in Medical portal different services. The Use Case describes the Actors as: Patient who uses website with Pre-condition having a valid email address to complete registration. A in Medical portal sends message to new doctor to activate his/her account. Also a message will be sent to A in Medical.com's Administrator (new Patient has been registered), AinMedical.com presents welcome page for New Patient and provide link to login his account. As indicated in Fig.(4).

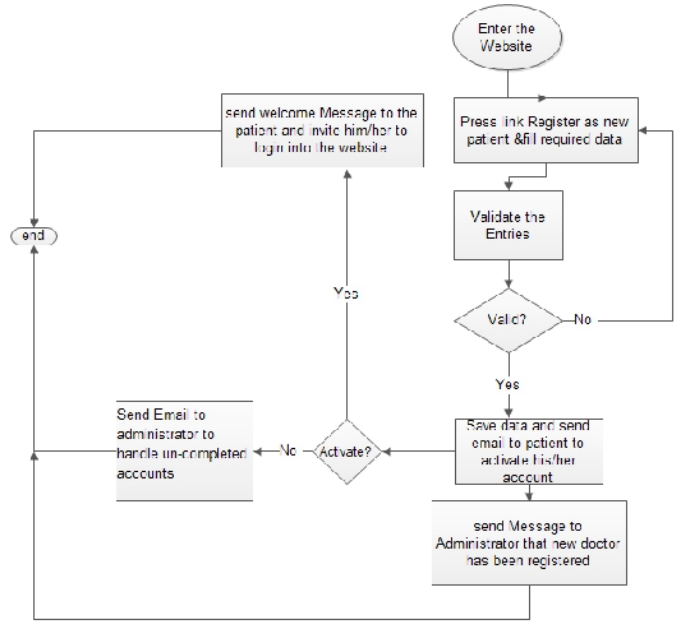

Figure 4. Patient Registration

**3) *Telemedicine services Activation*,** From Doctor side, the requirements of telemedicine services activation needs to manage doctor's account, the doctor already activated this service for his account then user press "Add/Manage Telemedicine service" link to activate it for the first time or manage his data in it respectively and the flow chart appeared in Fig.(5)

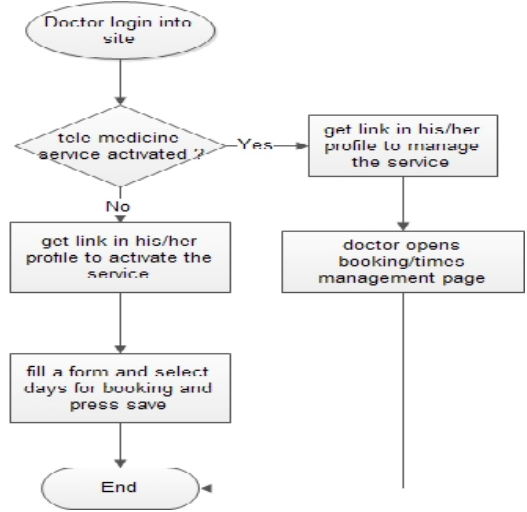

Figure 5. Telemedicine services Activation

**4) *Doctor Manage telemedicine services & their setting (days & time-slots /cost)*,** If Portal's doctor needs to review/mange his telemedicine services, it requires him to login on A in Medical Portal with his account then press "Manage Telemedicine service" link to review his telemedicine services, add new services and modify them. If Portal's doctor needs to review/mange telemedicine setting (days & time-slots /cost), it requires him to login on A in Medical Portal with his account then press "Manage Telemedicine service" link to review service setting, add new times and modify days & time-slots /cost.

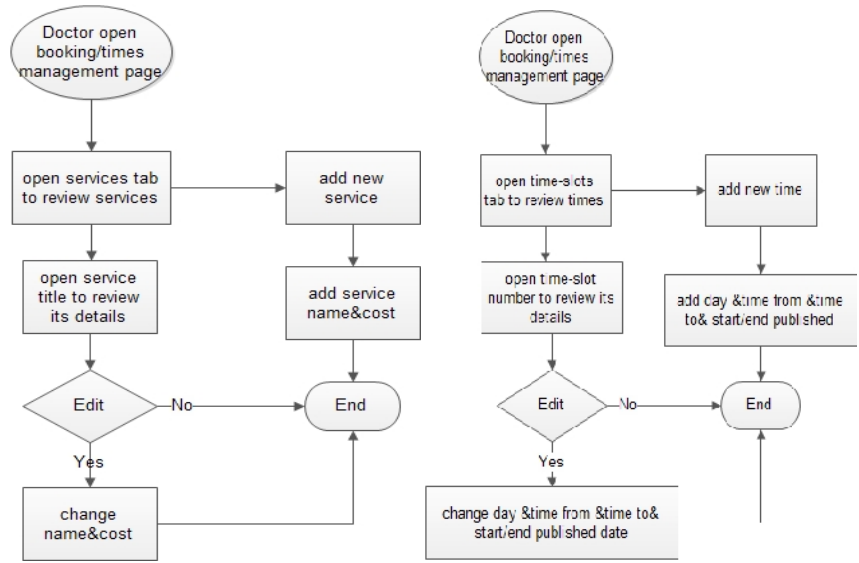

Figure 6. Doctor Manage telemedicine services & settings

**5)** ***Doctor Manage booking requests***, If Portal's doctor needs to needs to review/mange booking requests), it requires him to login on A in Medical Portal with his account then press "Manage Telemedicine service" to review booking requests and then the doctor can select request to review patient's details (patient data, time, patient medical history), the doctor can change request status or add prescription and required radiograph & tests.

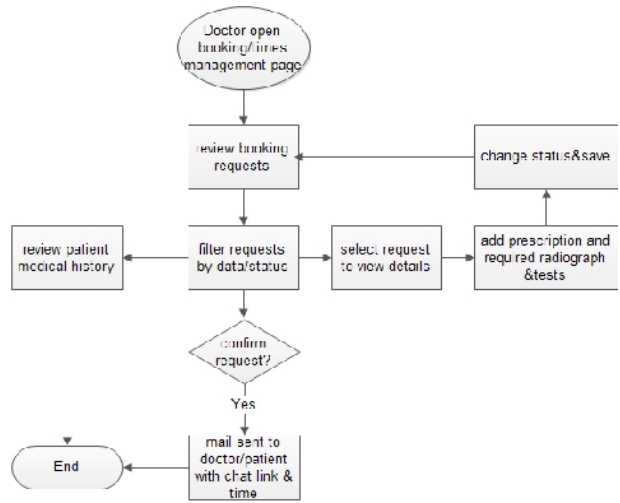

Figure 7.Doctor Manage booking requests

**6)***Patient add basic information & medical history***, From patient side, the requirements of telemedicine services activation needs to manage doctor's account, Portal's patient need to add his basic information. The patient enter to Telemedicine link on A in Medical Portal, then press " Add/View medical history", and select the needed link to review/add data for (basic information, diseases, symptoms ,pharmaceuticals, surgeries, sensitivities, radiograph, tests).

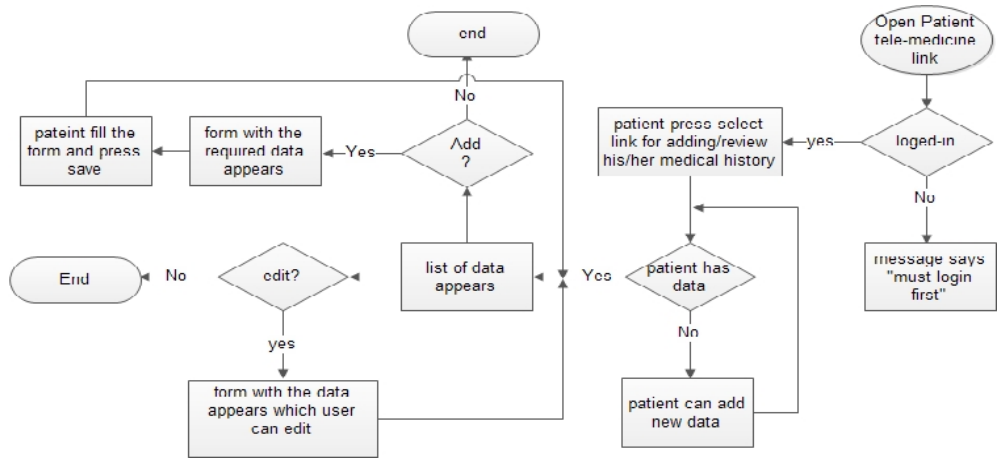

Figure 8.Patient add basic information & medical history

**7)** ***Patient makes booking &reviews his booking list***, If patient needs to use telemedicine services and make booking with one of AinMedical.com doctors, this requires the patient to review doctor list to search for the required doctor. The patient press "book now" which beside the doctor he selects, fills the form of booking then selects the time and press send. The request will be sent and the payment taken from patient credit.

If patient needs to review his booking list. the patient enter to Telemedicine link on AinMedical.com  Portal, then user press "booking data" link & review his booking list. If the conversation with doctor done, patient can review the prescription and the required radiographs& tests.

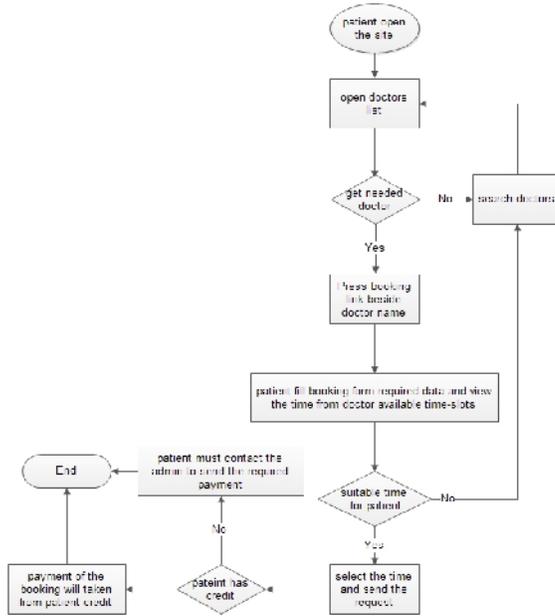
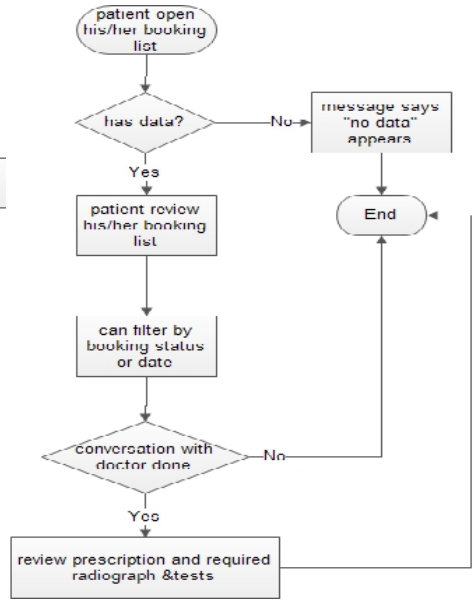

Figure 9. Patient makes booking          Figure 10.  Patient reviews his booking list

## 3. CONCLUSIONS

This paper describes a telemedicine system used to provide medical services to a rural healthcare center. Three operational modes of the telemedicine system are explored through the system developed. In order to fulfill the requirements of medical practice, we define a PIP that functions as a database processing element encapsulating medical information obtained during one patient visit. A PIP-based data structure can reduce the complexity of accessing medical information. In this study, we also integrate multimedia patient information within the same database system and provide two kinds of user interfaces for different medical service purposes.

The medical services provided by the telemedicine system at the rural site are eagerly needed by the elderly. The system allows the elderly to avoid traveling a long distance to get better care. Evaluation results show that the telemedicine system is relatively feasible in the case of teleradiology. Telemedicine has shown the capability not only to improve the quality of healthcare, but also to increase the opportunity of continuing education for physicians at a rural site. According to the results of the survey, the WWW environment's features of multimedia and hyper linking made the web-based browser suitable for displaying medical teaching materials.

Based on the system developed, there are many other aspects that can be explored in the future. One is to add a datamining technique to the system [32]. This could allow the formulation of diagnostic behaviors and build a knowledge base to assist diagnosis and medical teaching. The other is to incorporate image compression technique to speed image transmission [33]. These advances may help researchers to not only explore the knowledge of medical behavior, but also expand the feasibility of the telemedicine system.

# REFERENCES

[1]    B. H. Guze, R. Estep, and C. Fisher, "Telemedicine: A review of its use and a proposal for application in psychiatric consultation," Med. Inform., vol. 20, no. 1, pp. 1–18, 1995.

[2]    J. E. Cabral, Jr. and Y. Kim, "Multimedia systems for telemedicine and their communications requirements," IEEE Commun. Mag., July 1996, pp. 20–27.

[3]    T. Paakkala, J. Aalto, V. K¨ah¨ar¨a, and S. Sepp¨anen, "Diagnostic performance of a teleradiology system in primary health care," Comput. Methods Programs Biomed., vol. 36, pp. 157–160, 1991.

[4]    J. Viitanen, T. Sund, E. Rinde, J. Stoermer, M. Kormano, J. Heinila, J.Yliaho, and J. Ahonen, "Nordic teleradiology development," Comput.Methods Programs Biomed., vol. 37, pp. 273–277, 1992.

[5]    H. K. Huang, "Teleradiology technologies and some service models," Comput. Med. Imag. Graph., vol. 20, no. 2, pp. 59–68, 1996.

[6]    O. Ratib, Y. Ligier, and J. R. Scherrer, "Digital image management and communication in medicine," Comput. Med. Imag. Graph., vol. 18, no.2, pp. 73–84, 1994.

[7]    H. K. Huang et al., "Implementation of a large-scale picture archiving and communication system," Comput. Med. Imag. Graph., vol. 17, no. 1, pp. 1–11, 1993.

[8]    D. F. Leotta and Y. Kim, "Requirements for picture archiving andcommunications," IEEE Eng. Med. Biol. Mag. , pp. 62–69, Mar. 1993.

[9]    H. K. Huang, W. K. Wong, S. L. Lou, and B. K. Stewart, "Architecture of a comprehensive radiologic imaging network," IEEE J. Select. Areas Commun., vol. 10, pp. 1188–1196, Sept. 1992.

[10]   W. J. Chimiak, "The digital radiology environment," IEEE J. Select.AreasCommun., vol. 10, pp. 1133–1144, Sept. 1992.

[11]   S. T. Treves, E. S. Hashem, B. A. Majmudar, K. Mitchell, and D. J. Michaud, "Multimedia communications in medical imaging," IEEE J. Select. Areas Commun., vol. 10, pp. 1121–1132, Sept. 1992.

[12]   S. L. Lou, J. Wang, M. Moskowitz, T. Bazzill, and H. K. Huang, "Methods of automatically acquiring images from digital medical systems," Comput. Med. Imag. Graph., vol. 19, no. 4, pp. 369–376, 1995.

[13]   Digital Imaging and Communications in Medicine (DICOM) Version 3.0, Amer. College Radiologists/Nat. Elect. Manufacturers Assoc., 1993.

[14]   G. Bucci, R. Detti, S. Nativi, and V. Pasqui "Loosely coupled workstations in a radiological image information system," Future Generation Comput. Syst., vol. 8, pp. 31–42, 1992.

[15]   S. K. Mun, M. Freedman, and R. Kapur, "Image Management and communications for radiology," IEEE Eng. Med. Biol. Mag., pp. 70–80, Mar. 1993.

[16]   G. Hartviksen, S. Akselsen, A. K. Eidsvik, S. Pedersen, and E. Rinde, "Toward a general purpose, scaleable workstation for remote medical consultations.Experiences from use of VIDA-a still image system for the provision of low-cost telemedicine," Med. Inform., vol. 20, no. 1, pp. 19–33, 1995.

[17]   H. Handels, C. Busch, J. Encarna¸cao C. Hahn, V. K hn, J. Miehe, S. I. P¨oppl, E. Rinast, C. Roßmanith, F. Seibert, and A. Will, "KAMEDIN: A telemedicine system for computer supported cooperative work and remote image analysis in radiology," Comput. Methods Programs Biomed., vol. 52, pp. 175–183, 1997.

[18]   F. R. Bartsch, M. Gerneth, and R. Schosser, "Videoconference as a tool for European inter-hospital consultations in radiology," in Proc. SPIE,1977, pp. 62–67.

[19]   S. J. Dwyer, III, et al., "Teleradiology using switched dialup networks," IEEE J. Select. Areas Commun., vol. 10, pp. 1161–1172, Sept. 1992.

[20]   L. Orozco-Barbosa, A. Karmouch, N. D. Georganas, and M. Goldberg, "A multimedia interhospital communications system for medical consultations," IEEE J. Select. Areas Commun., vol. 10, pp. 1145–1157, Sept. 1992.

[21]   K. Chipman, P. Holzworth, J. Loop, N. Ransom, D. Spears, and B. Thompson, "Medical applications in a B-ISDN field trial," IEEE J. Select. Areas Commun., vol. 10, pp. 1173–1187, Sept. 1992.

[22] H. K. Huang, R. L. Arenson, S.-L. Lou, A. W. K. Wong, K. P. Andriole, T. M. Bazzill, and D. Avrin, "Multimedia in the radiology environment:Current concept," Comput. Med. Imag. Graph., vol. 18, no. 1, pp. 1–10, 1994.

[23] G. F. Egan and Z.-Q.Liu, "Computers and networks in medical and healthcare systems," Comput.Biol. Med., vol. 25, no. 3, pp. 355–365, 1995.

[24] T. Kitanosono, Y. Kurashita, M. Honda, T. Hishida, H. Konishi, M. Mizuno, and M. Anzai, "The use of multimedia in patient care," Comput. Methods Programs Biomed., vol. 37, pp. 259–263, 1992.

[25] S. T. C. Wong and H. K. Huang "A hospital integrated framework for multimodality image base management," IEEE Trans. Syst., Man, Cybern. A, vol. 26, pp. 455–469, July 1996.

[26] R. K. Taira, B. K. Stewart, and U. Sinha, "PACS database architecture and design," Comput. Med. Imag. Graph., vol. 15, no. 3, pp. 171–176, 1991.

[27] S. Badaoui, V. Chameroy, and F. Aubry, "A database manager of biomedical images," Med. Inform., vol. 18, no. 1, pp. 23–33, 1993.

[28] F. Pinciroli, C. Combi, and G. Pozzi, "ARCADIA: A system for the integration of angiocardiographic data and images by an object-oriented DBMS," Comput. Biomed.Res., vol. 28, pp. 5–23, 1995.

[29] A. Karmouch, "Multimedia distributed cooperative system," Comput. Commun., vol. 16, pp. 568–580, Sept. 1993.

[30] R. E. Rakel, Textbook of Family Practice, 5th ed. Philadelphia, PA: Saunders, 1995.

[31] Digital Imaging and Communications in Medicine (DICOM) Version 3.0, Amer. College Radiologists/Nat. Elect. Manufacturers Assoc., 1993.

[32] C. T. Liu, C. C. Lin, J. M. Wong, S. K. Chiou, R. S. Chen, J. H. Chen, S. M. Hou, and T. Y. Tai, "Design and evaluation of a telediagnosis system," Biomed. Eng. Applicat., Basis Commun., vol. 9, pp. 52–60, Apr. 1997.

[33] H. S. Chen et al., "Integrated medical informatics with small group teaching in medical education," Int. J. Med. Inform., to be published.

## Author

Hafez Fouadreceivedhis BSc. degree in Electronics and communications engineering in 1993, EGYPT and received his M.Sc. and Ph.D. degrees from Ain Shams University in 2001 and 2008. His Ph.D. is dedicated to Performance Optimization of CMOS RF Power Amplifiers for Mobile Communication systems. The M.Sc. is dedicated to Design and Optimization of Silicon RF Front-ends For Mobile Communication Systems. He is a researcher at the Electronics Research Institute (ERI), Ministry of Scientific Research, Cairo, Egypt. His current research interests are Telemedicine Systems, wireless sensors network, Bioelectronics, Bioinformatics and their applications.

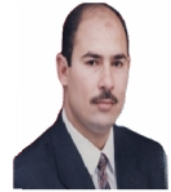